\documentclass[preprint,3p,12pt]{elsarticle}

\usepackage{amsmath,amssymb,amsfonts}
\usepackage{bm,bbm}
\usepackage[utf8]{inputenc}
\usepackage[T1]{fontenc}
\usepackage{booktabs}
\usepackage{url}
\usepackage{hyperref}

\newcommand{\durdevich}{{\DJ}ur{\dj}evich}

\newcommand{\nwse}[3]{\ensuremath{#1^{#2}_{\phantom{#2} #3}}}
\newcommand{\osp}{\ensuremath{\mathfrak{osp} \left( 32 \middle\vert 1 \right)}}

\journal{Nuclear Physics B}

\begin{document}

\begin{frontmatter}

\title{On eleven-dimensional Supergravity and Chern--Simons theory}

\author[ucsc,unam,valencia]{Fernando Izaurieta}
\ead{fizaurie@ucsc.cl}

\author[ucsc]{Eduardo Rodr\'{\i}guez}
\ead{edurodriguez@ucsc.cl}

\address[ucsc]{Departamento de Matemática y Física Aplicadas, Universidad Católica de la Santísima Concepción\\
Alonso de Ribera 2850, 4090541 Concepción, Chile}

\address[unam]{Instituto de Matemáticas, Universidad Nacional Autónoma de México\\
Av.\ Insurgentes Sur s/n, México D.F.}

\address[valencia]{Departament de Física Teòrica, Universitat de València\\
C/ Dr.\ Moliner 50, 46100 Burjassot, Valencia, Spain}

\begin{abstract}
We probe in some depth into the structure of eleven-dimensional, \osp-based Chern--Simons supergravity,
as put forward by Troncoso and Zanelli (TZ) in 1997.
We find that the TZ Lagrangian may be cast as a polynomial in $1/l$, where $l$ is a length,
and compute explicitly the first three dominant terms.
The term proportional to $1/l^{9}$ turns out to be essentially the Lagrangian of the standard 1978 supergravity theory of Cremmer, Julia and Scherk, thus establishing a previously unknown relation between the two theories.
The computation is nontrivial because, when written in a sufficiently explicit way,
the TZ Lagrangian has roughly one thousand non-explicitly Lorentz-covariant terms.
Specially designed algebraic techniques are used to accomplish the results.
\end{abstract}

\begin{keyword}
Non-abelian gauge theories \sep Chern--Simons theories \sep Supersymmetry and Supergravity
\end{keyword}

\end{frontmatter}

\section{Introduction}
\label{sec:intro}

Eleven-dimensional, \osp-based Chern--Simons (CS) Supergravity was put forward by
R.~Troncoso and J.~Zanelli (TZ) in 1997~\cite{Tro97,Tro98}.
The appearance of this eleven-dimen\-sional theory followed previous efforts in lower dimensions by
S.~Deser, R.~Jackiw and S.~Templeton~\cite{Des82b,Des83},
A.~Achúcarro and P.~K.~Townsend~\cite{Ach86}, and
A.~H.~Chamseddine~\cite{Cha90}.

The ``supergravity'' name has proven a bit controversial.
On one hand, the TZ theory includes gravity and fermionic matter and sports an exact, off-shell
\osp\ symmetry which mixes bosons and fermions.
On the other, it differs quite radically%
\footnote{For instance, the TZ theory lacks equality between bosonic and fermionic degrees of freedom (at least in the off-shell counting).
This matching, which has been taken as a watermark of supersymmetry in the literature, lies also at the foundations of the MSSM.
The recent negative results of the search for supersymmetric particles at the LHC~\cite{CMS11a,Str11,ATL11a,ATL11b,Buc11},
although clearly still at an early stage,
should not then be taken as an irrefutable sign that supersymmetry as a principle is wrong.}
from the ``standard'' 1978 theory of Cremmer, Julia and Scherk (CJS)~\cite{Cre78},
which is widely regarded as unique~\cite{Des97}.

The purpose of this paper is to shed new light on the relation between the TZ and the CJS theories.

In this regard, we would like to highlight here the work of
M.~Bañados~\cite{Banh01}, who showed that the linearized perturbations of the TZ theory are the same as the ones of CJS supergravity
(see also Ref.~\cite{Hor97}).
That result, however, is based on the identification of the CJS three-form $A_{3}$ with $e^{a} T_{a}$
(where $e^{a}$ and $T^{a}$ are the elfbein and torsion, respectively),
which might lead to inconsistencies~\cite{Ede08}.
Our own approach
suggests that
$A_{3}$ should to be related to the CS fields through a much more complicated ansatz
[see eq.~(\ref{eq:A3})].

Rather than focusing on the dynamics of the TZ theory, in this paper we deal with its Lagrangian structure.
This proves to be a more tractable approach, since CS theories possess highly nonlinear dynamics and a complicated phase space structure~\cite{Mis03}.

We first show that the TZ Lagrangian may be cast as a polynomial of degree~11 in $1/l$, where $l$ is a length.
When the CS five-index one-form $b^{abcde}$ vanishes, the leading term is just a cosmological constant,
while the sub-leading term includes both torsion and a mass term for the fermions.
Our main results follows:
the next term (proportional to $1/l^{9}$) is essentially the CJS Lagrangian (see section~\ref{sec:main}).
This discovery shows the existence of a previously unknown relation between the TZ and the CJS theories.

That this result may be new owes much to the great complexity of the TZ Lagrangian (see section~\ref{sec:TZ}).
While a compact closed formula exists [see eq.~(\ref{eq:TZLag})],
a sufficiently explicit version would amount to roughly a thousand terms and has never been attempted.

To achieve our result we have first split the TZ Lagrangian into several meaningful terms
by means of the CS subspace separation method introduced in Ref.~\cite{Iza06a}.
A careful dimensional analysis has then allowed us to extract only those terms proportional to the required powers of $l$.
For the sake of simplicity, we have restricted ourselves to the case when $b^{abcde} = 0$, which is roughly equivalent to setting $A_{3} = 0$.
The details of the calculation are summarized in~\ref{sec:CJSTZ}.

Further comments and an outlook for future work are given in section~\ref{sec:final}.

\section{A brief look at the TZ theory}
\label{sec:TZ}

The TZ theory is a pure gauge theory in eleven-dimen\-sional spacetime
whose Lagrangian $\mathcal{L}_{\text{TZ}}$ is the CS eleven-form for the \osp\ superalgebra.
A compact closed formula for $\mathcal{L}_{\text{TZ}}$ reads%
\footnote{Wedge product between differential forms is assumed throughout, e.g., $\bm{A}^{2} = \bm{A} \wedge \bm{A}$.}
\begin{equation}
 \mathcal{L}_{\text{TZ}} =
 6 \int_{0}^{1} dt 
 \left\langle
  \bm{A}
  \left(
   t \mathrm{d} \bm{A} +
   t^{2} \bm{A}^{2}
  \right)^{5}
 \right\rangle,
 \label{eq:TZLag}
\end{equation}
where $\bm{A}$ is an \osp-valued one-form and
$\left\langle \cdots \right\rangle$
stands for an \osp-invariant supersymmetric polynomial of rank six.%
\footnote{This is essentially a higher-order version of the Killing metric, which may include a number of dimensionless coupling constants.}

The general properties of CS forms~\cite{Che74,deAz95} ensure that $\mathcal{L}_{\text{TZ}}$
changes at most by a closed form under \osp\ gauge transformations, which close off-shell
without introducing auxiliary fields.

Choosing a convenient basis for \osp\ one may write $\bm{A}$ as
\begin{equation}
 \bm{A} =
 \frac{1}{l} e^{a} \bm{P}_{a} +
 \frac{1}{2} \omega^{ab} \bm{J}_{ab} +
 \frac{1}{5!l} b^{abcde} \bm{Z}_{abcde} +
 \frac{1}{\sqrt{l}} \bar{\bm{Q}} \psi,
 \label{eq:A}
\end{equation}
where
$\bm{P}_{a}$ are (AdS nonabelian) translations,
$\bm{J}_{ab}$ are Lorentz rotations,
$\bm{Z}_{abcde}$ are extra bosonic generators (not central charges---$\bm{Z}_{abcde}$ behaves as a five-index, fully antisymmetric Lorentz tensor), and
$\bm{Q}$ is a fermionic generator.
This basis, with its associated commutation relations,
allows the interpretation of
$e^{a}$ as the orthonormal frame of one-forms%
\footnote{The spacetime metric $g_{\mu \nu}$ is not a fundamental field in the TZ theory,
being derived from $e^{a}$ by means of the relation
$g_{\mu \nu} = \eta_{ab} \nwse{e}{a}{\mu} \nwse{e}{b}{\nu}$.}
(the \textit{elfbein} one-form) and
$\omega^{ab}$ as the spin connection,
which shows that gravity (in its first-order formulation) is included in the TZ theory.

Dimensional consistency demands that we introduce a length parameter $l$ in eq.~(\ref{eq:A}).
This length scale allows the one-form $\bm{A}$ to remain dimensionless while giving the right dimensions to the component fields
$e^{a}$, $\omega^{ab}$, $b^{abcde}$ and $\psi$.

The definition of $\mathcal{L}_{\text{TZ}}$
is only complete after we specify the precise meaning of
$\left\langle \cdots \right\rangle$.
Perhaps the simplest way to build this invariant polynomial
is to use the trace of the supersymmetrized product of six generators in a suitable supermatrix representation.%
\footnote{This is not the most general choice, though, and more general invariant polynomials may prove useful. We return to this in section~\ref{sec:final}.}
Our choice for such supermatrix representation is as follows:
\begin{align}
\bm{P}_{a} & = \left[
\begin{array}[c]{cc}
\frac{1}{2} \nwse{\left( \Gamma_{a} \right)}{\alpha}{\beta} & 0 \\
0 & 0
\end{array}
\right], \\
\bm{J}_{ab}  &  =\left[
\begin{array}[c]{cc}
\frac{1}{2} \nwse{\left( \Gamma_{ab} \right)}{\alpha}{\beta} & 0 \\
0 & 0
\end{array}
\right], \\
\bm{Z}_{abcde}  &  =\left[
\begin{array}[c]{cc}
\frac{1}{2} \nwse{\left( \Gamma_{abcde} \right)}{\alpha}{\beta} & 0 \\
0 & 0
\end{array}
\right], \\
\bm{Q}^{\rho} & = \left[
\begin{array}[c]{cc}
0 & C^{\alpha \rho} \\
\delta_{\beta}^{\rho} & 0
\end{array}
\right],
\end{align}
where $\Gamma_{a}$, $\Gamma_{ab}$ and $\Gamma_{abcde}$ are Dirac Gamma matrices in $d=11$,
and $C_{\alpha \beta} = -C_{\beta \alpha}$ is the charge conjugation matrix.

The computation of
$\left\langle \cdots \right\rangle$
is thus reduced to the calculation of traces of products of Gamma matrices,
which is conceptually straightforward but may be quite challenging in practice.
In fact, the invariant polynomial so-defined has yet to be fully computed explicitly,
which means that the detailed structure of the TZ theory is not well known.

\section{CJS Supergravity and the TZ Theory}
\label{sec:main}

The inclusion of the length parameter $l$ in eq.~(\ref{eq:A}) and the algebraic structure of CS Lagrangians
[see eq.~(\ref{eq:TZLag})]
imply that the TZ Lagrangian can be written as a polynomial of degree~11 in $1/l$, namely
\begin{equation}
\mathcal{L}_{\text{TZ}} =
\sum_{p=0}^{11}
l^{-p}
\mathcal{L}^{\left( p \right)}.
\label{eq:LTZpol}
\end{equation}
We expect $l$ to be
a small quantity.
The most relevant terms in eq.~(\ref{eq:LTZpol}) are then
\begin{equation}
\mathcal{L}_{\text{TZ}} =
l^{-11} \mathcal{L}^{\left( 11 \right)} +
l^{-10} \mathcal{L}^{\left( 10 \right)} +
l^{-9} \mathcal{L}^{\left( 9 \right)} +
\cdots.
\label{eq:LTZ3}
\end{equation}
Without yet evaluating the invariant polynomial
$\left\langle \cdots \right\rangle$,
the first two terms are given by
(see~\ref{sec:CJSTZ})
\begin{align}
\mathcal{L}^{\left( 11 \right)} & =
 \frac{6}{11}
 \left\langle
   \bm{e}
   \left(
     \bm{e}^{2}
   \right)^{5}
 \right\rangle, \label{eq:L11sim} \\
\mathcal{L}^{\left( 10 \right)} & =
 15
 \left\langle
  \bm{\psi}
  \left[ \bm{e} , \bm{\psi} \right]
  \left(
   \bm{e}^{2}
  \right)^{4}
 \right\rangle +
 3
 \left\langle
  \bm{eT}
  \left( \bm{e}^{2} \right)^{4}
 \right\rangle, \label{eq:L10sim}
\end{align}
where, for simplicity, we have set $\bm{b} = 0$.%
\footnote{Here we use the abbreviations
$\bm{e} = e^{a} \bm{P}_{a}$,
$\bm{\omega} = \frac{1}{2} \omega^{ab} \bm{J}_{ab}$,
$\bm{b} = \frac{1}{5!} b^{abcde} \bm{Z}_{abcde}$, and
$\bm{\psi} = \bar{\bm{Q}} \psi$.
All powers of $l$ are displayed explicitly.}
As advertised, the leading term is a cosmological constant,
while the sub-leading term includes both torsion and a mass term for the fermions.

The evaluation of the next term
(the one proportional to $1/l^{9}$)
is significantly more involved,
since it requires the computation of many components of
$\left\langle \cdots \right\rangle$.
We have left all details concerning this calculation for~\ref{sec:CJSTZ},
where we also explain the notation and conventions used below.
The result reads
\begin{align}
\mathcal{L}^{\left( 9 \right)} & =
- \frac{1}{4 \times 9! \kappa_{2}^{2}} \varepsilon_{a_{1} \cdots a_{11}} R^{a_{1} a_{2}} e^{a_{3}} \cdots e^{a_{11}}
- \frac{7i}{4} \bar{\psi} \Gamma_{\left( 8 \right)} \mathrm{D} \psi
- \frac{1}{2} ab
+ \nonumber \\ &
+ \frac{i}{8} \left( T^{a} - \frac{i \kappa_{1}^{2}}{4} \bar{\psi} \Gamma^{a} \psi \right) e_{a} \bar{\psi} \Gamma_{\left( 6 \right)} \psi
+ \frac{7}{2^{4} \times 5} a \left. \ast a \right.
- \frac{1}{2^{10} \times 3^{2}} \frac{\left( 4! \right)^{5}}{6!} b \left. \ast b \right.,
\label{eq:L9}
\end{align}
where we have also set $\bm{b} = 0$.

Compare now eq.~(\ref{eq:L9}) with the CJS Lagrangian~\cite{Jul99},
\begin{align}
\mathcal{L}_{\text{CJS}} & =
- \frac{1}{4 \times 9! \kappa^{2}} \varepsilon_{a_{1} \cdots a_{11}} R^{a_{1} a_{2}} e^{a_{3}} \cdots e^{a_{11}}
+ \frac{i}{2} \bar{\psi} \Gamma_{(8)} \mathrm{D} \psi
- \frac{1}{2} ab
+ \nonumber \\ &
+ \frac{i}{8} \left( T^{a} - \frac{i \kappa^{2}}{4} \bar{\psi} \Gamma^{a} \psi \right) e_{a} \bar{\psi} \Gamma_{\left( 6 \right)} \psi
- \frac{1}{2} \left( a + F \right) \left. \ast F \right.,
\label{eq:LCJS}
\end{align}
where we have set $A_{3} = 0$.

The similarity between eqs.~(\ref{eq:L9}) and~(\ref{eq:LCJS}) is, of course, striking.
Apart from some slightly different numerical coefficients,
the main difference is to be found among the last few terms,
which are precisely those more sensitive to the truncation made by setting
$\bm{b} = 0$ and $A_{3} = 0$.

It could be argued that this similarity is just accidental,
and that, given the highly nonlinear structure of the TZ Lagrangian,
almost any imaginable kind of term will appear in it.
Interestingly enough, this is not the case:
there are many possible combinations which could in
principle appear in the $l^{-9}$ sector of the TZ Lagrangian,
and which do not show up in the CJS supergravity Lagrangian.
None of these appear.
Some examples are
\begin{align}
\left[ EB \right]^{abcde} & = \nwse{E}{abcd}{f} B^{fe}, \label{noshowfirst} \\
\left[ EBBB \right]^{abcde} & = \nwse{E}{ab}{fgh} B^{fc} B^{gd} B^{he}, \\
\left[ EBB^{2} \right]^{abcde} & = \nwse{E}{abc}{fg} B^{fd} \nwse{B}{g}{h} B^{he}, \\
\left[ EB^{3} \right]^{abcde} & = \nwse{E}{abcd}{f} \nwse{B}{f}{g} \nwse{B}{g}{h} B^{he}, \\
\left[ BEB \right]^{abc} & = B_{de} \nwse{E}{deab}{f} B^{fc}, \\
\left[ BBEB \right]^{a} & = B_{bc} B_{de} \nwse{E}{bcde}{f} B^{fa}, \label{noshowlast}
\end{align}
where $B^{ab}$ is any antisymmetric two-index tensor constructed from the physical fields
(e.g., $e^{a} e^{b}$, $R^{ab}$, etc.).
Similarly, $E^{abcde}$ stands for any antisymmetric five-index combination, for which,
even with $b^{abcde}=0$, there are several possibilities:
$\bar{\psi} \Gamma^{abcde} \psi$, $e^{a} e^{b} e^{c} e^{d} e^{e} $, etc.


\section{Discussion}
\label{sec:final}

We have shown that the TZ Lagrangian may be cast as a polynomial of degree 11 in $1/l$,
where $l$ is a (small) length parameter,
and that the term proportional to $1/l^{9}$
is essentially identical to the CJS Lagrangian.

One significance of our result lies on its improbability:
the TZ theory and CJS supergravity emerge from entirely different backgrounds,
yet both seem to be related at a deep level.
We would like to emphasize this:
to the best of our knowledge,
there is a priori no reason whatsoever for the TZ theory to be related in this way to CJS supergravity,
yet the relation appears.

For instance, the CJS Lagrangian includes the Hodge $\ast$-operator,
whereas the TZ Lagrangian, being formulated with no a priori background metric, does not.
This sole fact could be na\"{\i}vely argued to provide a reason for the absence of a relation between both theories.
However, terms such as $a \left. \ast a \right.$ do end up appearing in the TZ Lagrangian.
This can be traced back to the fact that a duality relation exists between different Dirac Gamma matrices in eleven-dimensional spacetime.

A further interesting aspect deals with the \osp\ symmetry:
our results suggest that the CJS Lagrangian fails to be invariant under \osp\
because it lacks all the other terms that ensure the invariance of the TZ Lagrangian.%
\footnote{It has been known for some time, however, that CJS supergravity has on-shell symmetry under a one-parameter family of superalgebras~\cite{DAu82,Ban04a,Ban04b}.}

The cosmological constant issue is seen here under a new light.
While it is well known that CJS supergravity forbids a cosmological constant,
the TZ theory includes it on a different level, namely, in the term proportional to $1/l^{11}$
rather than in the one proportional to $1/l^{9}$ (where the CJS terms reside).
The implications of this fact have yet to be fully assessed.

These observations lead us to propose the conjecture that CJS supergravity might actually be a truncation of a more general theory, such as the TZ theory.
In fact, the possibility that the CJS theory may be included within the TZ theory was already put forward by its authors~\cite{Tro97}.

Our result also highlights the usefulness of the CS subspace separation method~\cite{Iza06a}
for manipulating CS Lagrangians in higher-dimensional spacetime.
Given the complexities involved, such powerful methods become essential tools
to extract any meaningful information from CS theories (see~\ref{sec:CJSTZ}).

There are several ways in which this work could be extended.
The inclusion of a nonzero $b^{abcde}$ is obviously one of them,
but the technical hurdles involved may be hard to overcome.
Since only one-forms are allowed in CS theories,
the $A_{3}$ three-form must be first decomposed in terms of the TZ fields before a comparison is attempted.
The idea of decomposing $A_{3}$ in terms of one-forms has been already analyzed in the literature
(see, e.g., Refs.~\cite{Zan05,Ban04b}).
Our construction has, however, an additional constraint;
$A_{3}$ must be of order $l^{-3}$,
and therefore, there are forbidden combinations.
For instance, the gravitino cannot be part of $A_{3}$,
because it will lead to orders $l^{-1}$ and $l^{-2}$.
Derivatives of the fields are also forbidden,
since it would lead at most to order $l^{-2}$ and order $l^{-3}$ would never be reached.
Therefore, the identification of $A_{3}$ with a term such as $e^{a} T_{a}$ \cite{Banh01}
is ruled out from the outset.
A moment's thought shows that the only consistent way of decomposing $A_{3}$
in terms of the TZ fields is through the ansatz
\begin{align}
A_{3} & =
C_{abc} e^{a} e^{b} e^{c} +
\frac{1}{5!} C_{abc_{1} \cdots c_{5}} e^{a} e^{b} b^{c_{1} \cdots c_{5}}
+ \frac{1}{\left( 5! \right)^{2}} C_{ab_{1} \cdots b_{5} c_{1} \cdots c_{5}} e^{a} b^{b_{1} \cdots b_{5}} b^{c_{1} \cdots c_{5}} +
\nonumber \\ & +
\frac{1}{\left( 5! \right)^{3}} C_{a_{1} \cdots a_{5} b_{1} \cdots b_{5} c_{1} \cdots c_{5}} b^{a_{1} \cdots a_{5}} b^{b_{1} \cdots b_{5}} b^{c_{1} \cdots c_{5}},
\label{eq:A3}
\end{align}
where the various $C$-tensors are fully anti-symmetric zero-forms to be determined.

As mentioned in section~\ref{sec:TZ}, the invariant polynomial we have used is but the simplest choice.
More general invariant polynomials may prove useful for eliminating the slight differences that still remain between the CJS Lagrangian and the $1/l^{9}$ sector of the TZ theory.

Another interesting issue is the exploration of the solution space of the TZ theory.
While the full theory is difficult to analyze due to the nonlinearities of the Lagrangian~\cite{Banh95,Banh96b,Cha99,Mis03},
the truncated version (at order $1/l^{9}$) should prove to be more tractable.

\section*{Acknowledgments}

The authors wish to thank
José~A.\ de~Azcárraga,
Alfredo Pérez,
Patricio Salgado and
Jorge Zanelli
for many fruitful discussions regarding the topics covered in the present work.
We are also indebted to
Ricardo Troncoso
for helpful comments on the first draft of this manuscript.
F.~I.\ extends his thanks to
José~A.\ de~Azcárraga
for his kind hospitality at the Department of Theoretical Physics of the Universitat de Val\`{e}ncia, and to
Micho \durdevich\
for his kind hospitality at the Institute of Mathematics of the Universidad Nacional Autónoma de México, where part of the present work was done.
E.~R.\ is grateful to
Jorge Zanelli
for his kind hospitality at the Centro de Estudios Científicos (CECS), Valdivia, where part of this work was done.
The authors also thank
Kasper Peeters
for his computer algebra system ``Cadabra'' \cite{Pee06,Pee07}, which was of invaluable help when performing lengthy Dirac-matrix algebra computations, and
Ricardo Ramírez,
for helping to create the necessary code to evaluate some of the components of the invariant polynomial.
F.~I.\ and E.~R.\ were supported by the
National Commission for Scientific and Technological Research, Chile, through Fondecyt research grants 11080200 and 11080156, respectively.

\appendix

\section{Detailed calculation of the first three orders of the TZ Lagrangian}
\label{sec:CJSTZ}

\subsection{The TZ Lagrangian and its complexities}

Na\"{\i}ve use of eq.~(\ref{eq:TZLag}) results in an explicit expression for the TZ Lagrangian that contains around a thousand non-explicitly Lorentz-covariant terms.
However, in order to compare with the CJS Lagrangian, it is necessary to have each term expressed using Lorentz-covariant quantities,
such as the Lorentz curvature $R^{ab}$ and torsion $T^{a}$.
In principle, this requires several very carefully-chosen integrations by parts for each one of these thousand terms.

To avoid such a nightmarish calculation, we have followed a different route,
one that makes use of custom algebraic techniques.

First, we use transgression forms to systematically write the TZ Lagrangian in terms of Lorentz-covariant quantities.
This is what we call the ``CS subspace separation method,'' introduced in Refs.~\cite{Iza06a,Iza05}.

We then cast the TZ Lagrangian as a polynomial in $1/l$, where $l$ is a length [see eq.~(\ref{eq:A})].
This is relevant because the CJS Lagrangian can be related only to the terms in the $l^{-9}$ sector~\cite{DAu82}.

As a last step, we explicitly evaluate the invariant polynomial
$\left\langle \cdots \right\rangle$
[see eq.~(\ref{eq:TZLag})],
a calculation which involves heavy use of Dirac matrices and their properties.
While conceptually straightforward, this computation is quite challenging in practice, requiring the use of computer algebra software~\cite{Pee06,Pee07,maxima} and new numerical techniques~\cite{Iza11b}.

\subsection{Expressing the TZ Lagrangian in terms of Lorentz-covariant quantities}

As explained fully in Ref.~\cite{Iza06a}, any CS Lagrangian can be split as a sum of terms that reflect the subspace structure of the gauge algebra.
For the case of the TZ Lagrangian~(\ref{eq:TZLag}) we find
\begin{equation}
\mathcal{L}_{\text{TZ}} =
\mathcal{L}_{\bm{\psi}} +
\mathcal{L}_{\bm{b}} +
\mathcal{L}_{\bm{e}} +
\mathcal{L}_{\bm{\omega}} +
\mathrm{d} \mathcal{B}.
\label{eq:LTZssm}
\end{equation}
The last term in this sum is an exact form which can be easily calculated using the techniques from Ref.~\cite{Iza06a},
but which shall not interest us here, since it amounts to a boundary term in the action.
The remaining four terms are examples of
\textit{transgression forms}~\cite{Iza05,Bor03,Mor04a,Bor05,Mor06a,Mor06b}.

A transgression form
$Q^{\left( 2n+1 \right)} \left( \bm{A} ; \bar{\bm{A}} \right)$
is defined as the following function of the two one-form gauge connections $\bm{A}$ and $\bar{\bm{A}}$~\cite{deAz95}:
\begin{equation}
Q^{\left( 2n+1 \right)} \left( \bm{A} ; \bar{\bm{A}} \right) =
\left( n+1 \right) \int_{0}^{1} dt \left\langle
\left( \bm{A} - \bar{\bm{A}} \right) \bm{F}_{t}^{n} \right\rangle,
\label{eq:deftrans}
\end{equation}
where $\bm{F}_{t}$ is the curvature associated with the interpolating connection
$\bm{A}_{t} = \bar{\bm{A}} + t \left( \bm{A} - \bar{\bm{A}} \right)$,
$\bm{F}_{t} = \mathrm{d} \bm{A}_{t} + \bm{A}_{t}^{2}$.


Using the definition~(\ref{eq:deftrans}) we find that we can write the first four terms in eq.~(\ref{eq:LTZssm}) as follows:
\begin{align}
\mathcal{L}_{\bm{\psi}} & =
Q^{\left( 11 \right)} \left(
\frac{1}{l} \bm{e} + \frac{1}{l} \bm{b} + \frac{1}{\sqrt{l}} \bm{\psi} + \bm{\omega}
;
\frac{1}{l} \bm{e} + \frac{1}{l} \bm{b} + \bm{\omega}\right), \\
\mathcal{L}_{\bm{b}} & =
Q^{\left( 11 \right)} \left(
\frac{1}{l} \bm{e} + \frac{1}{l} \bm{b} + \bm{\omega}
;
\frac{1}{l} \bm{e} + \bm{\omega} \right), \\
\mathcal{L}_{\bm{e}} & =
Q^{\left( 11 \right)} \left( \frac{1}{l} \bm{e} + \bm{\omega}
;
\bm{\omega} \right), \\
\mathcal{L}_{\bm{\omega}} & =
Q^{\left( 11 \right)} \left(
\bm{\omega} ; 0 \right).
\end{align}
All fermionic terms have now been packaged in
$\mathcal{L}_{\bm{\psi}}$,
while the other three are purely bosonic.
In particular,
$\mathcal{L}_{\bm{e}}$
depends only on the elfbein and the spin connection, and thus corresponds to the gravitational sector.

While not immediately obvious from the above expressions,
splitting a CS Lagrangian in terms of transgression forms plus a boundary term provides a very elegant mechanism for writing the TZ Lagrangian in terms of Lorentz-covariant quantities (exception made of the ``exotic'' gravity term $\mathcal{L}_{\bm{\omega}}$).

\subsection{Casting the TZ Lagrangian as a Polynomial in $1/l$}

As eq.~(\ref{eq:A}) shows, the one-form $\bm{A}$ includes different powers of the length parameter $l$.
Let us label by
$\bm{A}^{\left( p \right)}$ and $\bm{F}^{\left( p \right)}$
the terms of order $l^{-p/2}$ of the connection and curvature.
We then have
\begin{align}
\bm{A} & =
 \bm{A}^{\left( 0 \right)} +
 \bm{A}^{\left( 1 \right)} +
 \bm{A}^{\left( 2 \right)}, \\
\bm{F} & =
 \bm{F}^{\left( 0 \right)} +
 \bm{F}^{\left( 1 \right)} +
 \bm{F}^{\left( 2 \right)} +
 \bm{F}^{\left( 3 \right)} +
 \bm{F}^{\left( 4 \right)},
\label{eq:F01234}
\end{align}
since $\bm{A}$ contains only the powers $0$, $-1/2$ and $-1$ of $l$
and $\bm{F}$ is at most quadratic in $\bm{A}$.
Every term in~(\ref{eq:F01234}) can now be written as%
\footnote{These partial curvatures are best computed using a generalized version of the Gauss--Codazzi equations.
Let $\bm{A}$ and $\bar{\bm{A}}$ be two one-form connections,
and let $\bm{\Delta} \equiv \bm{A} - \bar{\bm{A}}$.
The curvatures $\bm{F}$ and $\bar{\bm{F}}$ are related by the identity
$\bm{F} = \bar{\bm{F}} + \bar{\mathrm{D}} \bm{\Delta} + \bm{\Delta}^{2}$,
where $\bar{\mathrm{D}}$ stands for the covariant derivative in the connection $\bar{\bm{A}}$.}
\begin{align}
\bm{F}^{\left( 0 \right)} & = \bm{R}, \label{eq:F0} \\
\bm{F}^{\left( 1 \right)} & = \frac{1}{l^{1/2}} \mathrm{D}_{\omega} \bm{\psi}, \\
\bm{F}^{\left( 2 \right)} & = \frac{1}{l} \left( \bm{T} + \mathrm{D}_{\omega} \bm{b} + \bm{\psi}^{2} \right), \\
\bm{F}^{\left( 3 \right)} & = \frac{1}{l^{3/2}} \left[ \bm{e} + \bm{b} , \bm{\psi} \right], \\
\bm{F}^{\left( 4 \right)} & = \frac{1}{l^{2}} \left( \bm{e}^{2} + \left[ \bm{e} , \bm{b} \right] + \bm{b}^{2} \right), \label{eq:F4}
\end{align}
where
$\bm{R} = \frac{1}{2} R^{ab} \bm{J}_{ab}$ and
$\bm{T} = T^{a} \bm{P}_{a}$
are the usual Lorentz curvature and torsion, and
$\mathrm{D}_{\omega}$
stands for the Lorentz covariant derivative.

Collecting the different powers of $l$ is now an exercise in combinatorics,
which can be performed without too much difficulty.

For $\mathcal{L}^{\left( 11 \right)}$ and $\mathcal{L}^{\left( 10 \right)}$ we find
\begin{align}
\mathcal{L}^{\left( 11 \right)}_{\bm{\psi}} & =
0, \label{eq:LTZ11psi} \\
\mathcal{L}^{\left( 11 \right)}_{\bm{b}} & =
 6 \int_{0}^{1} dt
 \left\langle
   \bm{b}
   \left(
     \bm{e}^{2} + t
     \left[
       \bm{e} , \bm{b}
     \right]
     + t^{2} \bm{b}^{2}
   \right)^{5}
 \right\rangle, \\
\mathcal{L}^{\left( 11 \right)}_{\bm{e}} & =
 \frac{6}{11}
 \left\langle
   \bm{e}
   \left(
     \bm{e}^{2}
   \right)^{5}
 \right\rangle, \\
\mathcal{L}^{\left( 11 \right)}_{\bm{\omega}} & = 0,
\end{align}
\begin{align}
\mathcal{L}^{\left( 10 \right)}_{\bm{\psi}} & =
 15
 \left\langle
  \bm{\psi}
  \left[ \bm{e} + \bm{b} , \bm{\psi} \right]
  \left(
   \bm{e}^{2} +
   \left[ \bm{e} , \bm{b} \right] +
   \bm{b}^{2}
  \right)^{4}
 \right\rangle, \\
\mathcal{L}^{\left( 10 \right)}_{\bm{b}} & =
 30 \int_{0}^{1} dt
 \left\langle
  \bm{b}
  \left( \bm{T} + t \mathrm{D}_{\omega} \bm{b} \right)
  \left(
   \bm{e}^{2} + t
   \left[ \bm{e} , \bm{b} \right] +
   t^{2} \bm{b}^{2}
  \right)^{4}
 \right\rangle, \\
\mathcal{L}^{\left( 10 \right)}_{\bm{e}} & =
 3
 \left\langle
  \bm{eT}
  \left( \bm{e}^{2} \right)^{4}
 \right\rangle, \\
\mathcal{L}^{\left( 10 \right)}_{\bm{\omega}} & = 0.
\end{align}

The result for $\mathcal{L}^{\left( 9 \right)}$ is significantly more complicated:
\begin{align}
\mathcal{L}^{\left( 9 \right)}_{\bm{\psi}} & =
\frac{6!}{4!} \frac{1}{2}
\left\langle
  \bm{\psi} \mathrm{D}_{\omega} \bm{\psi}
  \left(
    \bm{e}^{2} +
    \left[
      \bm{e}, \bm{b}
    \right]
    + \bm{b}^{2}
  \right)^{4}
\right\rangle +
\nonumber \\ & +
\frac{6!}{3!} \frac{1}{2}
\left\langle
  \bm{\psi}
  \left(
    \bm{T} + \mathrm{D}_{\omega} \bm{b} + \frac{2}{3} \bm{\psi}^{2}
  \right)
  \left[
    \bm{e} + \bm{b}, \bm{\psi}
  \right]
  \left(
    \bm{e}^{2} +
    \left[
      \bm{e}, \bm{b}
    \right]
    + \bm{b}^{2}
  \right)^{3}
\right\rangle +
\nonumber \\ & +
\frac{6!}{3!2!} \frac{1}{4}
\left\langle
  \bm{\psi}
  \left[
    \bm{e} + \bm{b}, \bm{\psi}
  \right]^{3}
  \left(
    \bm{e}^{2} +
    \left[
      \bm{e}, \bm{b}
    \right]
    + \bm{b}^{2}
  \right)^{2}
\right\rangle, \label{eq:LTZ9phi}
\end{align}
\begin{align}
\mathcal{L}^{\left( 9 \right)}_{\bm{b}} & =
\frac{6!}{4!} \int_{0}^{1} dt
\left\langle
  \bm{bR}
  \left(
    \bm{e}^{2} + t
    \left[
      \bm{e}, \bm{b}
    \right]
    + t^{2} \bm{b}^{2}
  \right)^{4}
\right\rangle +
\nonumber \\ & +
\frac{6!}{2!3!} \int_{0}^{1} dt
\left\langle
  \bm{b}
  \left(
    \bm{T} + t \mathrm{D}_{\omega} \bm{b}
  \right)^{2}
  \left(
    \bm{e}^{2} + t
    \left[
      \bm{e}, \bm{b}
    \right]
    + t^{2} \bm{b}^{2}
  \right)^{3}
\right\rangle, \label{eq:LTZ9b}
\end{align}
\begin{align}
\mathcal{L}^{\left( 9 \right)}_{\bm{e}} & =
\frac{6!}{4!} \frac{1}{9}
\left\langle
 \bm{eR}
 \left( \bm{e}^{2} \right)^{4}
\right\rangle +
\frac{6!}{2!3!} \frac{1}{9}
\left\langle
 \bm{eT}^{2}
 \left( \bm{e}^{2} \right)^{3}
\right\rangle, \label{eq:LTZ9e} \\
\mathcal{L}^{\left( 9 \right)}_{\bm{\omega}} & = 0. \label{eq:LTZ9w} 
\end{align}


A quick glance at eqs.~(\ref{eq:LTZ11psi})--(\ref{eq:LTZ9w}) shows that the successive terms of the TZ Lagrangian increase rapidly in complexity.
To keep calculations at a manageable level, we shall restrict the following analysis to the case when $\bm{b} = 0$.
In this case we can write [cf.~eqs.~(\ref{eq:L11sim})--(\ref{eq:L10sim})]
\begin{align}
\mathcal{L}^{\left( 11 \right)} & =
 \frac{6}{11}
 \left\langle
   \bm{e}
   \left(
     \bm{e}^{2}
   \right)^{5}
 \right\rangle, \\
\mathcal{L}^{\left( 10 \right)} & =
 15
 \left\langle
  \bm{\psi}
  \left[ \bm{e} , \bm{\psi} \right]
  \left(
   \bm{e}^{2}
  \right)^{4}
 \right\rangle +
 3
 \left\langle
  \bm{eT}
  \left( \bm{e}^{2} \right)^{4}
 \right\rangle,
\end{align}
and
\begin{align}
\mathcal{L}^{\left( 9 \right)} & =
 \frac{6!}{4!} \frac{1}{9}
 \left\langle
  \bm{eR}
  \left( \bm{e}^{2} \right)^{4}
 \right\rangle +
 \frac{6!}{2!3!} \frac{1}{9}
 \left\langle
  \bm{eT}^{2}
  \left( \bm{e}^{2} \right)^{3}
 \right\rangle +
 \frac{6!}{4!} \frac{1}{2}
 \left\langle
  \bm{\psi} \mathrm{D}_{\omega} \bm{\psi}
  \left( \bm{e}^{2} \right)^{4}
 \right\rangle +
 \nonumber \\ & +
 \frac{6!}{3!} \frac{1}{2}
 \left\langle
  \bm{\psi}
  \left( \bm{T} + \frac{2}{3} \bm{\psi}^{2} \right)
  \left[ \bm{e}, \bm{\psi} \right]
  \left( \bm{e}^{2} \right)^{3}
 \right\rangle +
 \frac{6!}{3!2!} \frac{1}{4}
 \left\langle
  \bm{\psi}
  \left[ \bm{e}, \bm{\psi} \right]^{3}
  \left( \bm{e}^{2} \right)^{2}
 \right\rangle.
\end{align}
As advertised, the leading term is a cosmological constant,
while the sub-leading term includes both torsion and a mass term for the fermions.
By explicitly evaluating the invariant polynomial
$\left\langle \cdots \right\rangle$,
we shall show below that the term proportional to $1/l^{9}$ is essentially the CJS Lagrangian.

\subsection{Including the Invariant Polynomial}

The computation of the trace of the symmetrized product of a large number of Dirac matrices is conceptually straightforward but quite challenging in practice.
In this regard we have greatly benefited from use of the computer algebra system ``Cadabra''~\cite{Pee06,Pee07}, which allows powerful symbolic manipulation of Gamma matrices.

For the Lorentz subspace of \osp, a symmetric invariant polynomial can be computed from the algorithm put forward in Ref.~\cite{Iza11b},
which allows an efficient computation of the trace of the symmetrized product of an arbitrary number of Dirac matrices with two indices.
Potential generalizations of this algorithm may permit a more efficient calculation for symmetric invariant polynomials for more general algebras.

We find
\begin{align}
\mathcal{L}^{\left( 9 \right)} & =
  \frac{1}{3584}
  \varepsilon_{a_{1} \cdots a_{11}}
  R^{a_{1} a_{2}}
  \left( e^{a_{3}} \cdots e^{a_{11}} \right)
  - 945
  \left( \bar{\psi} \Gamma_{\left( 8 \right)} \mathrm{D}_{\omega} \psi \right)
  + \frac{135}{2}
  \left[ e^{c} T_{c} - \frac{1}{3} \left( \bar{\psi} \Gamma_{\left( 1 \right)} \psi \right) \right]
  \left( \bar{\psi} \Gamma_{\left( 6 \right)} \psi \right) +
\nonumber \\ & -
  \frac{75}{4}
  \left( \bar{\psi} \Gamma_{\left( 2 \right)} \psi \right)
  \left( \bar{\psi} \Gamma_{\left( 5 \right)} \psi \right)
  - \frac{7}{61440}
  \varepsilon_{a_{1} \cdots a_{11}}
  \left( e_{a_{12}} e_{a_{13}} e_{a_{14}} \right)
  \left( \bar{\psi} \Gamma^{a_{1} \cdots a_{5}} \psi \right)
  \left( {e}^{a_{6}} \cdots e^{a_{9}} \right)
  \left( \bar{\psi} \Gamma^{a_{10} \cdots a_{14}} \psi \right) +
\nonumber \\ & +
  \frac{1}{1024}
  \varepsilon_{a_{1} \cdots a_{11}}
  \left( \bar{\psi} \Gamma^{a_{1} a_{2}} \psi \right)
  \left( \bar{\psi} \Gamma^{a_{3} a_{4}} \psi \right)
  \left( e^{a_{5}} \cdots e^{a_{11}} \right).
\end{align}

Consider now the rescaling of the gauge connection $\bm{A}$ to
\begin{equation}
\bm{A} =
\frac{1}{l} e^{a} \bm{P}_{a} +
\frac{1}{2} \omega^{ab} \bm{J}_{ab} +
\frac{1}{5!l} b^{abcde} \bm{Z}_{abcde} +
\frac{3}{2} \frac{1+i}{\sqrt{2l}} \bar{\bm{Q}} \psi,
\end{equation}
and observe that the following identities are satisfied%
\footnote{The fact that terms such as these, which involve the Hodge $\ast$-operator, can appear in CS Lagrangians is quite unexpected.}
\begin{align}
a \left. \ast a \right. & = -
  \frac{3 \times 5}{2^{6}7!}
  \varepsilon_{a_{1} \cdots a_{11}}
  \left( \bar{\psi} \Gamma^{a_{1} a_{2}} \psi \right)
  \left( \bar{\psi} \Gamma^{a_{3} a_{4}} \psi \right)
  e^{a_{5}} \cdots e^{a_{11}},
\\
b \left. \ast b \right. & = -
  \frac{7!}{2^{6} 4! 4! 4! 5! 5!}
  \varepsilon_{a_{1} \cdots a_{11}}
  e_{a_{12}} e_{a_{13}} e_{a_{14}}
  \left( \bar{\psi} \Gamma^{a_{1} \cdots a_{5}} \psi \right)
  e^{a_{6}} \cdots e^{a_{9}}
  \left( \bar{\psi} \Gamma^{a_{10} \cdots a_{14}} \psi \right),
\end{align}
with $a$ and $b$ chosen as
\begin{align}
a & = \frac{i}{4} \sqrt{\frac{5}{2}} \left( \bar{\psi} \Gamma_{\left( 2 \right)} \psi \right), \\
b & = \frac{i}{4} \sqrt{\frac{5}{2}} \left( \bar{\psi} \Gamma_{\left( 5 \right)} \psi \right).
\end{align}
As a result, the $1/l^{9}$ term of the TZ Lagrangian can be rewritten as
\begin{align}
\mathcal{L}^{\left( 9 \right)} & =
- \frac{1}{4 \times 9! \kappa_{2}^{2}} \varepsilon_{a_{1} \cdots a_{11}} R^{a_{1} a_{2}} e^{a_{3}} \cdots e^{a_{11}}
- \frac{7i}{4} \bar{\psi} \Gamma_{\left( 8 \right)} \mathrm{D}_{\omega} \psi
- \frac{1}{2} ab
+ \nonumber \\ &
+ \frac{i}{8} \left( T^{a} - \frac{i \kappa_{1}^{2}}{4} \bar{\psi} \Gamma^{a} \psi \right) e_{a} \bar{\psi} \Gamma_{\left( 6 \right)} \psi
+ \frac{7}{2^{4} \times 5} a \left. \ast a \right.
- \frac{1}{2^{10} \times 3^{2}} \frac{\left( 4! \right)^{5}}{6!} b \left. \ast b \right.,
\label{eq:L9b}
\end{align}
where
$\kappa_{1}$ and $\kappa_{2}$ are numerical coefficients given by
\begin{align}
\kappa_{1} & = i \sqrt{3}, \\
\kappa_{2} & = i \frac{3}{20} \sqrt{\frac{5}{7}}.
\end{align}

The similarity between $\mathcal{L}^{\left( 9 \right)}$ and the CJS Lagrangian [cf.~eq.~(\ref{eq:LCJS})] is now in evidence,
especially when we observe that the last couple of terms are of the kind which would arise when
$F$ includes $a$ and $\ast b$ besides $\mathrm{d} A_{3}$.


\bibliographystyle{utphys}
\bibliography{biblio2011}

\end{document}